\documentstyle[11pt,graphicx,color,cancel,cite,ulem,framed,hyperref,url]{article}
\newcommand{\sq}   {\mbox{$\widetilde{q}$}}
\newcommand{\msq}  {\mbox{$M_{\sq}$}}
\newcommand{\tb}  {tan$\beta$}

\begin{document}


\begin{flushright}
{\tt IPMU14-0237}
\end{flushright}

\begin{center}
{\large \bf The role of MSSM heavy Higgs production in the self coupling measurement of 
the 125 GeV Higgs boson at the LHC }

\vskip 0.3cm
Biplob Bhattacherjee $^{a,b}$\footnote{Author's present address is $a$. Email: biplob@cts.iisc.ernet.in },
Arghya Choudhury$^{c}$\footnote{Email: arghyachoudhury@hri.res.in}, 
\vskip 0.3cm
{$^a$  Centre for High Energy Physics, 
Indian Institute of Science, Bangalore 560012, India}\\
{$^b$
Kavli IPMU(WPI), University of Tokyo, Kashiwa, Chiba-277-8583, Japan}

{$^c$
Regional Centre for Accelerator-based Particle Physics,\\
Harish-Chandra Research Institute, Allahabad - 211109, India} \\

\end{center}

\begin{abstract}
Measurement of the self coupling of 125 GeV Higgs boson is one of the most crucial tasks for  
high luminosity run of the LHC and it can only be measured in the di-Higgs final state. 
In the minimal supersymmetric standard model, heavy CP even Higgs ($H$) can decay into 
lighter 125 GeV Higgs boson ($h$) and therefore influence the di-Higgs production. 
We investigate the role of single $H$ production in the measurement of self coupling of $h$.
We find that $H \rightarrow hh$ decay can nontrivially affect the $h$ self coupling measurement   
in low tan$\beta$ regime when the mass of the heavy Higgs boson lies  between 250 - 600 GeV and depending 
on the parameter space it may be seen as an enhancement of the self coupling of 125 GeV Higgs boson. 
\end{abstract}

\newpage
\setcounter{footnote}{0}

\section{Introduction}
One of the long standing  problems of particle physics is the origin of mass of fundamental particles. 
In the standard model (SM), a scalar doublet is introduced, neutral component (called Higgs boson) of  
which spontaneously breaks the electroweak symmetry by acquiring the non-zero vacuum expectation value (vev) 
and consequently generates masses for all SM particles. The mass of the Higgs boson is a free parameter in 
the SM and it is determined by the vev of the Higgs field and Higgs quartic self coupling ($\lambda$). 
A new boson with a mass about 125-126 GeV has been recently observed  by 
ATLAS~\cite{Aad:2012tfa} and  CMS~\cite{Chatrchyan:2012ufa} collaborations of Large 
Hadron Collider (LHC) experiment which may be the only missing piece of the SM, i.e., the Higgs boson. 
The next crucial step is to measure the 
properties of the newly discovered boson and establish the connection between this particle and the electroweak 
symmetry breaking mechanism. The measured couplings of the new boson with fermions and gauge bosons are found 
to be quite compatible with those of the SM Higgs boson \cite{ATLAS-CONF-2014-009, CMS-PAS-HIG-14-009} 
and more accurate measurement will be performed in the near 
future at the 13/14 TeV LHC \cite{ATLAS-collaboration:1484890}. 
In order to reconstruct the full profile of the Higgs boson, we  also need to measure the Higgs 
self coupling along with other couplings. In the framework of SM, it is possible to determine Higgs self-coupling 
$\lambda$ from the accurate measurement of Higgs mass and vev of the Higgs field. However, we should note that 
this type of estimation is indirect in nature and independent confirmation is indeed required to prove the existence 
of SM Higgs boson.  The direct way to determine the coupling $\lambda$ is to produce three Higgs bosons through 
Higgs boson quartic coupling $\lambda$ in collider experiments. However, triple Higgs boson production cross 
section is too small to observe at the LHC even with very high luminosity and therefore the only probe is to 
observe di-Higgs production via Higgs trilinear coupling. Higgs trilinear coupling is generated by the electroweak 
symmetry breaking and it is proportional to $\lambda$ and the vev of the Higgs field. It is thus possible to measure 
the Higgs quartic self coupling $\lambda$ from the di-Higgs production cross section 
in the SM \cite{Boudjema:1995cb, Djouadi:1999rca,Plehn:1996wb}. The Higgs pair 
production cross section in the SM is also small (a few tens of fb at the 14 TeV LHC) and it is accessible at the very high 
luminosity LHC, called HL-LHC. \\

In 2015, the LHC will start to operate at 13/14 TeV center of mass energy and after 2018 it is expected that the LHC will 
be upgraded for high luminosity operation. At HL-LHC, prospect of SM Higgs self-coupling measurement has been 
studied extensively  
in Ref~\cite{Lafaye:2000ec, Baur:2003gp, Dolan:2012rv, Baglio:2012np, Goertz:2013kp, Yao:2013ika, Barger:2013jfa} 
using Higgs pair production process. Although,  we can have various final states 
like $b \bar b b \bar b$, $b \bar b W^+ W^-$, $b \bar b \gamma \gamma$, $b \bar b \tau  \bar \tau$, $W^{+}W^{-}W^{+}W^{-}$ etc.  
from Higgs pair production, phenomenological studies  show that $b \bar b \gamma \gamma$ channel is the most 
promising one~\cite{Baur:2003gp, Baglio:2012np}. 
A recent study \cite{ATLAS-collaboration:1484890} by ATLAS Collaboration has also confirmed the 
importance of $b \bar b \gamma \gamma$ 
channel for the measurement of SM Higgs self-coupling.  \\

Precise measurement of Higgs self-coupling and the 
reconstruction of Higgs potential is necessary in order to prove the correctness of the  SM as there are many well motivated 
scenarios beyond the standard model (BSM) which can have extended Higgs sector and/or non-standard 
couplings \cite{Dolan:2012ac, Chen:2013emb,Baglio:2014nea,Hespel:2014sla,Osland:1998hv}. The 
presence of new particles and couplings can potentially change the value of Higgs self coupling $\lambda$ compared 
to that of SM \cite{Dolan:2012ac, Chen:2013emb,Baglio:2014nea,Hespel:2014sla,Osland:1998hv}. 
In case, we observe deviation of $\lambda$ from its SM value, this can be the indication of new 
physics beyond the SM. \\
  
In minimal version of supersymmetric standard model (MSSM), which is one of the most favourable  BSM models, 
there are five Higgs bosons: one CP even light Higgs ($h$), one CP even heavy Higgs ($H$), one CP odd Higgs 
($A$) and two charged Higgs bosons ($H^\pm$).  In this scenario, the lightest CP even Higgs boson (h) can be identified 
with the observed 125 GeV Higgs boson  and other Higgs bosons may be discovered in future LHC run. In the 
MSSM scenario, tree level couplings of Higgs bosons depend on two parameters: Higgs mixing angle $\alpha$ 
and $\beta$, where $\tan \beta$ is the ratio of the vacuum expectation values of two Higgs doublets. This 
means that Higgs couplings can be significantly different from the SM depending on the parameter space of the model. \\
 
In MSSM, one of the important consequences of the presence of heavy Higgs boson is that $H$ can decay to lighter 
Higgs boson $h$ and in that case, single production of heavy Higgs can be seen as a pair production of lighter 
Higgs bosons. 
We already mentioned that the observation of pair production of Higgs boson is the  only direct 
way to measure the self coupling of $h$ and therefore, the production of heavy Higgs boson and its decay to $h$ can potentially 
 affect the measurement of $\lambda$.  In this paper, we study the effect of heavy Higgs ($H$) production  
on self coupling measurement of SM-like Higgs boson $h$ in the context of MSSM. We assume that the observed Higgs 
with mass 125 GeV is indeed MSSM lightest Higgs boson and the couplings of $h$ is such that it behaves like SM 
Higgs boson. The plan of the paper is as follows: in Sec.\ref{sec:br} we briefly discuss the production of 
heavy CP even MSSM Higgs and its decay to lighter Higgs boson.
 In Sec.\ref{sec:benchmark}, we introduce the 
benchmark points for further analysis and study the properties of 125 GeV Higgs in the light of 
LHC data. In Sec.\ref{sec:analysis} we illustrate how 
the production of $H$ can influence the self coupling measurement of lighter (SM like) Higgs boson. Summary 
of our work and possible issues are discussed in Sec.\ref{sec:conclusion}. 

\section{MSSM $H$ production and its decay to 125 GeV Higgs}
\label{sec:br}

 The couplings of the MSSM Higgs bosons are determined by two parameters: $\alpha$ and $\beta$ 
defined in the introduction. The couplings of gauge bosons with $h$ and $H$ are proportional to $\sin (\beta - \alpha)$ 
 and $\cos (\beta - \alpha)$  respectively. As the experimental data show that the observed Higgs boson, which is 
 assumed to be h,  couples to W/Z gauge boson similar to SM Higgs, $\sin (\beta - \alpha)$ should be close to unity 
in order to satisfy the above mentioned constraint. In the limit $\sin (\beta - \alpha) \rightarrow $1,  $h$ couples to fermions and gauge bosons exactly like SM Higgs boson. This is 
 quite similar to the case of decoupling limit in which the lightest MSSM Higgs boson behaves like SM Higgs boson. In this case, 
 heavier CP even Higgs $H$ does not couple to electroweak gauge bosons and coupling to up and down type fermions are either 
 suppressed or enhanced by $\tan \beta$. For large values of $\tan\beta$, the coupling of H (also $H^\pm$ and $A$) 
 to $b$ quark becomes strong as it scales with $ m_b \tan \beta$ while its coupling with the top quark, which is 
 $\propto m_t/ \tan \beta$, becomes rather weak. In that case, H dominantly decays to b quarks and $\tau$s. In other 
 words, $H \rightarrow h h$ branching is negligible for high value of $\tan \beta$. \\
 
\begin{figure}[!htb]
\begin{center}
{\includegraphics[angle =270, width=0.75\textwidth]{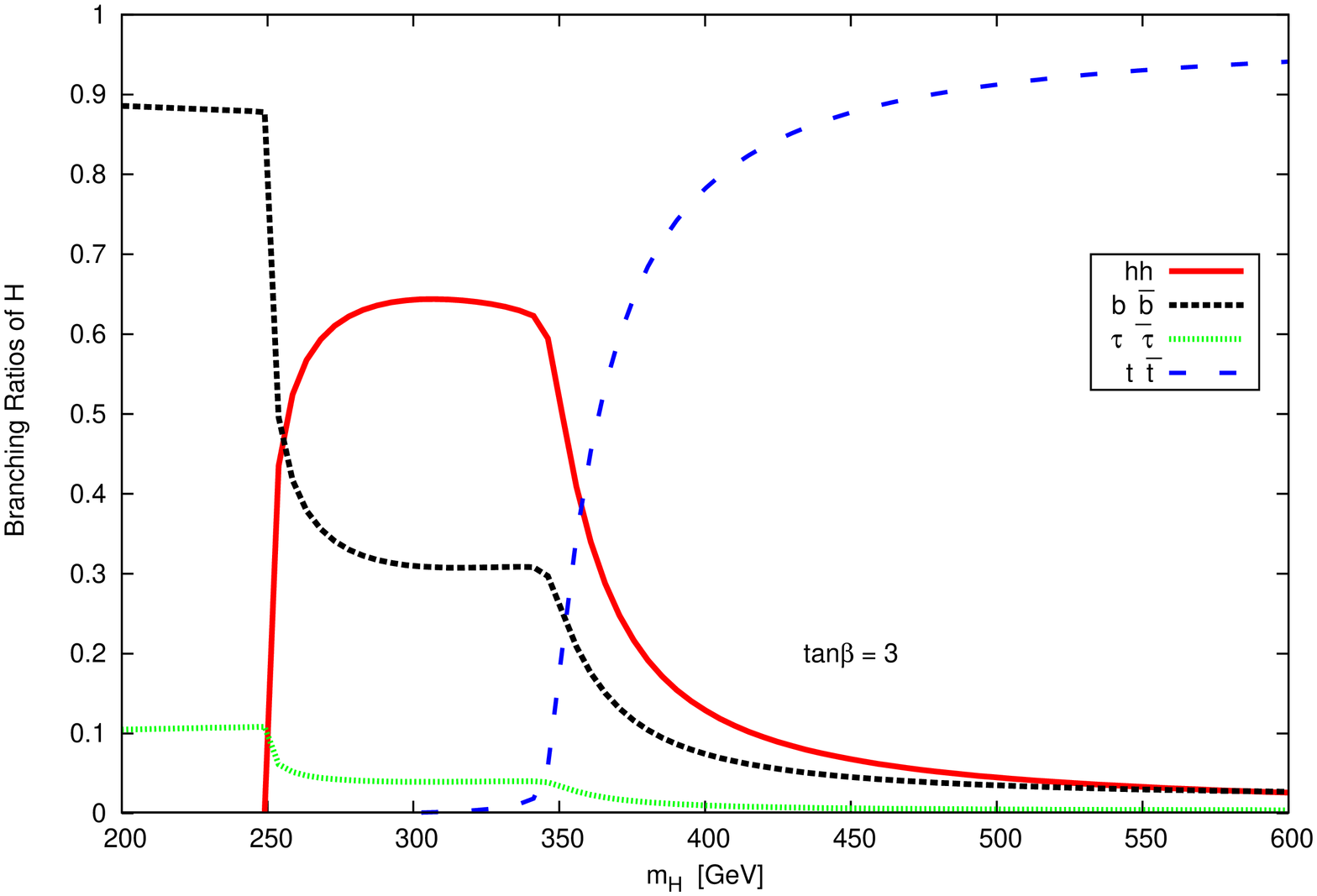} }
{\includegraphics[angle =270, width=0.75\textwidth]{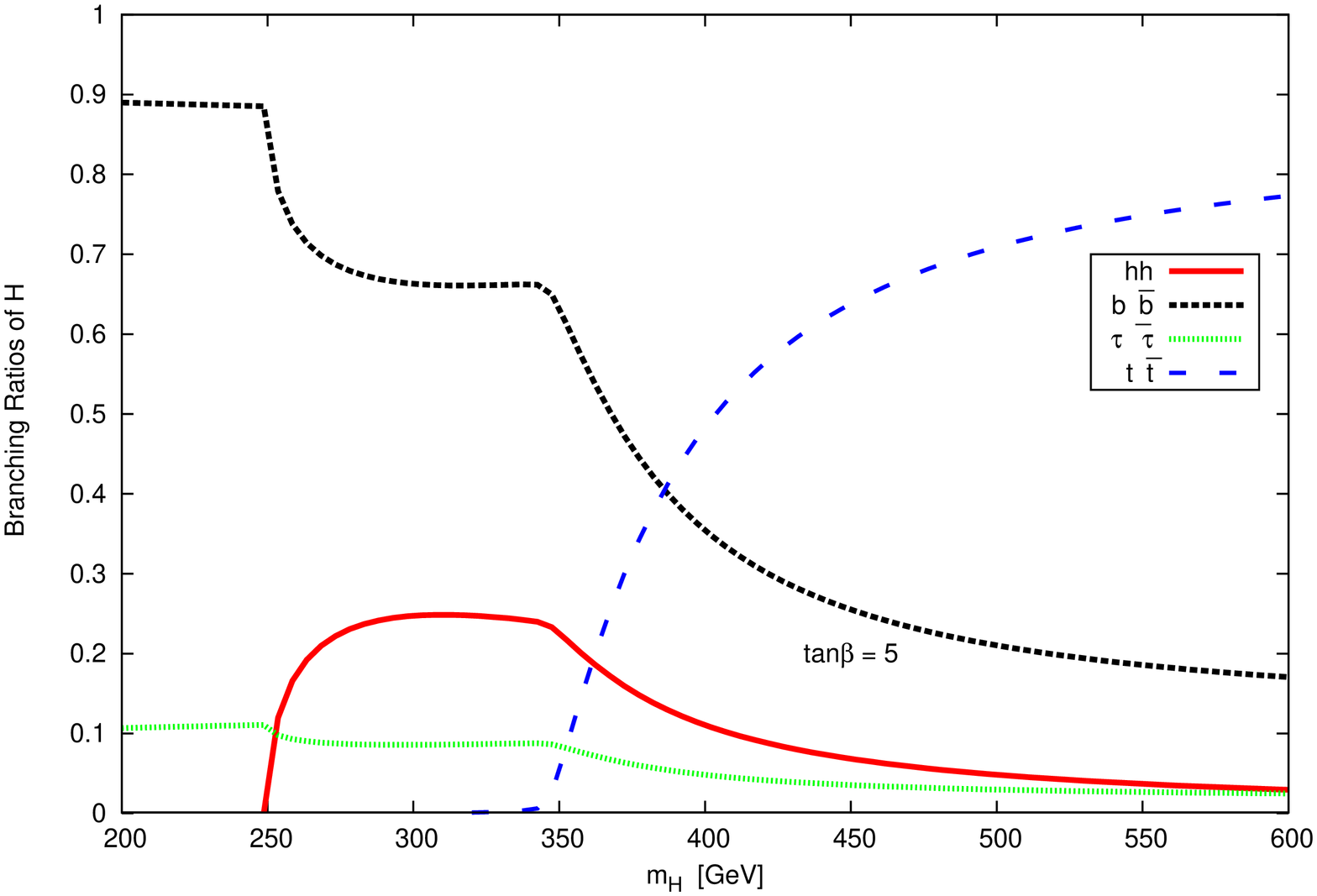} }
\caption{ {\it  Branching Ratios for heavy CP even Higgs ($H$) for $\tan\beta$ = 3 and 5. 
Results are given in the limit $\beta - \alpha = \pi/2$.  }}
\label{fig1}
\end{center}
\end{figure}


Here, we are interested to study the effect of heavy MSSM Higgs $H$ production 
which decays into a pair of $h$. 
In MSSM this decay mode is particularly important in the mass window 
$250$ GeV $< M_H < 400$ GeV for low value of $\tan \beta$. Above $M_{H}>350$ GeV, 
its decay to top quark opens up and eventually becomes the dominant one.  
To illustrate this point, we present the plot for branching ratio of $H$ 
for two values of $\tan\beta$ = 3, and 5 in Fig.~\ref{fig1} assuming 
$\sin(\beta - \alpha) = $1. For spectrum generation and 
decay/branching calculations, we use SUSY-HIT \cite{Djouadi:2006bz}. 
It is clear from Fig.~\ref{fig1} that for $\tan\beta$ = 3 (5) 
BR($H \rightarrow hh$) can be as high as 64 (25)\%. For $\tan\beta$ = 2, 
branching to lighter Higgs can go up to 83$\%$ while at moderate $\tan\beta$ (= 10) 
its maximum value reduces to only 2\%. For small $\tan\beta$, H branching to $t\bar t$ 
is important when it is kinematically allowed, but for large $\tan\beta$, BR($H \rightarrow b\bar b$) always 
dominates (for example, BR($H \rightarrow b\bar b$) varies between 70 to 90 \% for $\tan\beta$=10). 
In Table~\ref{tab1} we present the variation of branching ratio Br($H\rightarrow hh$) 
for different values of $M_H$ and $\tan\beta$ (see last three columns).  \\

\begin{table}[htbp]
\begin{center}\
\begin{tabular}{||c||c|c|c||c|c|c||}
\hline
$M_H$		& \multicolumn{3}{c||}{$\sigma_{NNLO}$ (pb)}& \multicolumn{3}{c||}{$Br(H\rightarrow hh)$} 	\\
\cline{2-4}
\cline{5-7}

(GeV)	& $\tan\beta$ =2  	&$\tan\beta$ = 3&$\tan\beta$ = 5  &$\tan\beta$ = 2	&$\tan\beta$ =3  &$\tan\beta$ = 5 \\
\hline
250	 	& 3.970		&2.416		& 2.459		&0.55		& 0.43		&0.12	\\
\hline
275	 	& 3.347		&1.945		& 1.798		&0.82		& 0.62		&0.22	\\
\hline
300	 	& 2.936		&1.635		& 1.371		&0.83		& 0.64		&0.25	\\
\hline
350	 	& 2.854		&1.460		& 0.959		&0.51		& 0.50		&0.22	\\
\hline
400	 	& 2.558		&1.233		& 0.676		&0.08		& 0.13		&0.11	\\
\hline
450	 	& 1.792		&0.844		& 0.431		&0.04		& 0.07		&0.07	\\
\hline
500	 	& 1.191		&0.554		& 0.273		&0.03		& 0.05		&0.05	\\
\hline
\hline

      \end{tabular}\
       \end{center}
           \caption{$\sigma_{NNLO}$ is the NNLO cross-section of single $H$ production from 
gluon fusion and bottom quark annihilation. 
Branching ratios of heavy CP even Higgs to di-Higgs final state ($Br(H\rightarrow hh)$) is calculated for different values of 
$\tan\beta$ assuming  $\beta - \alpha = \pi/2$ with $M_h = 125$ GeV. 
  }
\label{tab1}
          \end{table}

Production cross-section of $H$  depends on $\tan\beta$ through heavy quark couplings. 
The dominant contribution to single 
$H$ production mainly comes from gluon-gluon fusion although, for large or moderate 
$\tan\beta$, the bottom quark annihilation to $H$ 
($b \bar b \rightarrow H$) cross section can be substantial \cite{Han:2013sga}. We compute the NNLO 
cross-section of single $H$ production coming 
from gluon fusion and bottom-quark annihilation using SusHi (version 1.3.0) \cite{Harlander:2012pb} 
with MSTW 2008 (NNLO) PDF 
\cite{Martin:2009iq, Martin:2009bu, Martin:2010db}. The single production cross 
section of $H$ can be much larger than SM $hh$ production, 
for example, $\sigma(gg + b\bar b \rightarrow H)$ is 1635 $fb$ for $M_{H}$  = 300 GeV 
with $\tan\beta$ = 3. For the same $H$ mass, this cross section reduces to 1371 
$fb$ for $\tan\beta$ = 5. The total cross section $\sigma(gg+b\bar b \rightarrow H)$ for 
different benchmark points are presented in Table~\ref{tab1}. \\


\section{Benchmark points in pMSSM model}
\label{sec:benchmark}
We are interested to point out that the low $\tan\beta$ 
region in the MSSM is not well favoured due to the 
fact that in this region of the parameter space, it is very difficult to achieve the 
 lighter Higgs mass to $\sim$ 125 GeV assuming low supersymmetry (SUSY) breaking 
 scale ($M_S$). For example, with $M_S=$ 1 TeV and for $tan \beta$ = 3, the lightest 
 Higgs boson mass is just about $ \sim$ 99 GeV (assuming trilinear coupling $A_t$ = 0 GeV). 
However, it is possible to get 125 GeV Higgs mass  by lifting the SUSY 
scale to 5 - 100 TeV even with small $\tan \beta$ \cite{Degrassi:2012ry,Djouadi:2013vqa}. 
This type of high scale SUSY scenario has become particularly attractive in the context of recent 
discovery of 125 GeV Higgs boson which requires somewhat higher value of SUSY scale 
(typically  $M_S>$ 1 TeV) in contrast to the pre-LHC era, the absence of flavour changing 
neutral current and non observation of supersymmetric particles at the LHC. 
For these reasons, we define our benchmark points with large values of common scalar mass 
(a few TeV) for further analysis. We do not consider any particular form of SUSY breaking and we 
choose benchmark points in the phenomenological MSSM model (pMSSM). Below we talk about our choices 
of input parameters.

We set the values of gaugino mass parameters as $M_1$ = 1 TeV, $M_2$ = $M_3$ = 3  TeV. 
All squark and slepton masses (both L and R types) including third generations are assumed to be heavy 
and degenerate ( $M_{\tilde q}$ = $M_{\tilde q_L}$ = $M_{\tilde q_R}$ = 
$M_{\tilde l_L}$ = $M_{\tilde l_R}$). 
To achieve the Higgs mass we vary the common scaler mass $m_{\tilde q}$ 
(equivalent to $M_S$ or SUSY scale) from 3 TeV to 6 TeV. 
The other fixed parameters are $\mu$ = 1.5 TeV and the trilinear coupling $A_t$ = 0 (also the remaining
trilinear couplings $A_\tau,A_b,A_u,A_d,A_e$ parameters are all set to zero.)
In our analysis the relevant SM parameters considered are $m_t^{pole}=173.2$~GeV,
$m_b^{\overline {MS}}=4.19$~GeV and  $m_\tau=1.77$~GeV.


In Table \ref{tab2}, we present the relevant input parameters and output Higgs masses, 
Higgs mixing angle, sin$(\beta -\alpha)$ etc. for benchmark points BP1-BP5 
with $M_H$ = 275, 350, 450, 500 and 600 GeV respectively. 
From  Table \ref{tab2}, it is clear that sin$(\beta -\alpha)$ is very close 
to unity. 
The NNLO cross-section of single $H$ production ($\sigma_{NNLO}$), 
coming from gluon fusion and bottom-quark annihilation, has been presented in 
the last column of Table \ref{tab2}. We have computed $\sigma_{NNLO}$ 
using SusHi (version 1.3.0) \cite{Harlander:2012pb}.

\begin{table}[htbp]
\begin{center}\
\begin{tabular}{||c||c|c|c||c|c|c|c|c||c||}
\hline
		& \multicolumn{3}{c||}{Input Parameters}& \multicolumn{5}{c||}{$Output$}& 	\\

\cline{2-4}
\cline{5-9}
Point	&$\msq$	&$M_A$	&\tb	&$M_{h}$&$M_{H}$&$\alpha$ &sin$(\beta -\alpha)$&$Br(H$	&$\sigma_{NNLO}$ \\
	&(GeV)	&(GeV)	&	&(GeV)  &(GeV)	&	  &			&$\rightarrow hh)$&(pb)	\\
\hline
BP1	& 4350	&272	&5	&125.1	&275	&-0.2734  &0.9971		&28.2 	&2.265	\\
\hline
BP2	& 5820	&338	&2	&125.1	&350	&-0.5589  &0.9955		&45.6	&3.939\\
\hline
BP3	& 4125	&449	&6	&125.0	&450	&-0.1923  &0.9996		&6.1	&0.472\\
\hline
BP4	& 4275	&498	&5	&125.0	&500	&-0.2254  &0.9996		&5.1	&0.328\\
\hline
BP5	& 4275	&599	&5	&125.0	&600	&-0.2169  &0.9998		&2.2	&0.131\\

\hline
\hline

      \end{tabular}\
       \end{center}
           \caption{Input parameters and output masses, mixing angles, values of sin$(\beta -\alpha)$ 
for benchmark points BP1-BP5. $Br(H\rightarrow hh)$ represents  the branching ratios for heavy Higgs 
to light Higgs pair. $\sigma_{NNLO}$ is the NNLO cross-section of single $H$ production from 
gluon fusion and bottom quark annihilation.  }
\label{tab2}
          \end{table}


In the presence of mixing with the heavy CP even Higgs boson, the properties of the 125 GeV 
Higgs can change. It is thus important to check the branchings and production cross sections 
of 125 GeV Higgs in our case. Before presenting the analysis for heavy Higgs in the next section, now
we summarise the properties of the 125 GeV Higgs and MSSM effects on $hhh$ couplings for our 
selected benchmark points. 

Among the four main production modes of the Higgs boson at the LHC : 
gluon-gluon fusion (ggF), vector boson fusion (VBF), associated production with $W/Z$ 
bosons (VH) and top quarks ($t \bar t h$); ggF has the largest cross section and VBF 
process is the next dominant one. ATLAS and CMS collaborations have analysed the LHC 
data in five decay channels ($X \bar X$): $X \bar X$ = $\gamma \gamma, WW^*, ZZ^*, b\bar b, 
\tau^+ \tau^-$. They have also presented the signal strength ($\mu_{ggF/VBF/VH/t \bar t h}$) 
in individual/combine mode by measuring the $ggF/VBF/VH/t \bar t h$ rate with normalized by the 
SM predictions. In this work, we mainly focus on ggF and VBF production modes for illustration  
purpose.
\footnote{Both ATLAS and CMS group have measured signal strength in $b \bar b$ channel 
only from associated production with $W/Z$ bosons (VH) process.}.

\begin{table}[htbp]
\begin{center}\
\begin{tabular}{||c||c|c||c|c||}
\hline
		& \multicolumn{2}{c||}{ATLAS}& \multicolumn{2}{c||}{$CMS$} 	\\

\cline{2-3}
\cline{4-5}
$h\rightarrow X \bar X$	&$\mu_{ggF}$		&$\mu_{VBF}$		&$\mu_{ggF}$		&$\mu_{VBF}$	\\
\hline
$\gamma \gamma$		& 1.32$\pm$0.38\cite{atlas_gamma}&0.8$\pm$0.7\cite{atlas_gamma}	&$1.12^{+0.37}_{-0.32}$\cite{cms_gamma}	&$1.58^{+0.77}_{-0.68}$\cite{cms_gamma}		\\
\hline
$Z Z^*$			&$1.66^{+0.51}_{-0.44}$ \cite{atlas_zz}	&$0.26^{+1.64}_{-0.94}$\cite{atlas_zz}	&$0.80^{+0.46}_{-0.36}$\cite{cms_zz}	&$1.70^{+2.2}_{-2.1}$\cite{cms_zz}		\\
\hline
$W W^*$			&$1.02^{+0.29}_{-0.26}$	\cite{atlas_ww}&$1.27^{+0.53}_{-0.45}$\cite{atlas_ww}	&$0.74^{+0.22}_{-0.20}$	\cite{cms_ww}&$0.60^{+0.57}_{-0.46}$	\cite{cms_ww}	\\
\hline
$b \bar b$		& \multicolumn{2}{c||}{$0.51^{+0.40}_{-0.37}$ ($\mu_{VH}$)\cite{atlas_bb}}& \multicolumn{2}{c||}{$1.0 \pm 0.5$ ($\mu_{VH}$)\cite{cms_bb}} 	\\
\hline
$ \tau^+ \tau^-$	&$1.93^{+1.45}_{-1.15}$	\cite{atlas_tau}&$1.24^{+0.58}_{-0.54}$	\cite{atlas_tau}&$1.07 \pm 0.46$ \cite{cms_tau}	&$0.94 \pm 0.41$\cite{cms_tau}		\\
\hline
\hline
      \end{tabular}\
       \end{center}
           \caption{Latest results on signal strength ($\mu$) from LHC 7+8 TeV data  for the 
decay modes $h\rightarrow \gamma \gamma, WW^*, ZZ^*, b\bar b$ and  $\tau^+ \tau^-$. It may be noted that 
in the ZZ$^{*}$ channel, signal strength has been presented in 
two channels: combined $ggF+ b\bar b h + t \bar t h$ final states and combined $VBF +VH$ final states.}
\label{tab3}
          \end{table}

Signal strength in a particular channel $X\bar X$ is defined as 
\begin{equation}
\mu_{ggF/VBF}(X \bar X) = \frac{\Gamma(h\rightarrow gg/WW )}{\Gamma(h_{SM}\rightarrow gg/WW)}  
		      \times 
                      \frac {BR(h\rightarrow X \bar X)}{BR(h_{SM}\rightarrow X \bar X)} \nonumber
\end{equation}
where $\Gamma(h\rightarrow gg/WW)$ is the partial decay width used for $ggF/VBF$ mechanism.

We present the experimentally measured values of signal strength ($\mu$) from LHC 
7+8 TeV data with integrated luminosity 25 ${fb}^{-1}$ in Table \ref{tab3}. 
From the signal strength measurements one can constrain the couplings of Higgs with 
fermions or bosons. 
The measured $\mu$ values indicate that the 125 GeV Higgs is very close to SM like, 
but still there are several channels where the measured $\mu$ values are deviated from SM expectations 
and the errors are also large. 
We present the calculated $\mu$ values in ggF and VBF mode 
for our selected SUSY benchmark points in Table~\ref{tab4}. 
We can see from Table \ref{tab4} that in case of BP1, the signal strengths deviate appreciably 
from SM expectations although these values are within $2\sigma$ range of experimental 
results (see Table~\ref{tab3}). 
As we increase $M_A$ or $M_H$, the $\mu_{ggF/VBF}$ values become close to unity. 
All the benchmark points representing $M_H$ in the range 275 - 600 GeV, satisfy 
the latest data for 125 GeV Higgs within $2\sigma$ limits.

\begin{table}[htbp]
\begin{center}\
\begin{tabular}{||c||c|c||c|c||c|c||c|c||c|c||}
\hline
		&\multicolumn{2}{c||}{$BP1$}&\multicolumn{2}{c||}{$BP2$}& \multicolumn{2}{c||}{BP3}& \multicolumn{2}{c||}{$BP4$} &\multicolumn{2}{c||}{$BP5$}\\

\cline{2-3}
\cline{4-5}
\cline{6-7}
\cline{8-9}
\cline{10-11}
$h\rightarrow X \bar X$  &$\mu_{ggF}$&$\mu_{VBF}$&$\mu_{ggF}$&$\mu_{VBF}$&$\mu_{ggF}$&$\mu_{VBF}$&$\mu_{ggF}$&$\mu_{VBF}$&$\mu_{ggF}$&$\mu_{VBF}$\\
\hline
$\gamma \gamma$		&0.68 &0.71 &0.84 &0.90 &0.90	&0.90	&0.93 &0.93 &0.99 &0.98 	\\
\hline
$Z Z^*$			&0.67 &0.70 &0.82 &0.88 &0.89	&0.89	&0.92 &0.91 &0.98 &0.97 	\\
\hline
$W W^*$			&0.66 &0.69 &0.81 &0.87 &0.88	&0.88	&0.91 &0.91 &0.97 &0.96 	\\
\hline
$b \bar b$		&\multicolumn{2}{c||}{ 1.09 ($\mu_{VH}$)} &\multicolumn{2}{c||}{ 1.01 ($\mu_{VH}$)}&\multicolumn{2}{c||}{ 1.01 ($\mu_{VH}$)}&\multicolumn{2}{c||}{ 0.99 ($\mu_{VH}$)} &\multicolumn{2}{c||}{ 0.97 ($\mu_{VH}$)} 	\\
\hline
$ \tau^+ \tau^-$	&1.14 &1.20 &1.04 &1.12 &1.09	&1.09	&1.09 &1.08 &1.07 &1.06 	\\
\hline
\hline

      \end{tabular}\
       \end{center}
           \caption{ Signal strengths for the decay modes 
$h\rightarrow \gamma \gamma, WW^*, ZZ^*, b\bar b$ and  $\tau^+ \tau^-$ for benchmark points BP1-BP5. }
\label{tab4}
          \end{table}

\textbf{MSSM effects on hhh couplings:} 
Apart from the Higgs boson couplings to the gauge bosons and 
fermions, the trilinear and quartic couplings are also affected 
by the presence of MSSM. 
The trilinear couplings ($\lambda_{hhh}$) in SM and MSSM 
(with normalized to $\lambda_0 = [\sqrt{2} G_F]^{1/2} M_Z^2$) are given by \cite{Plehn:1996wb, Djouadi_rev}: 
\begin{eqnarray}
&SM\hphantom{SM}:& \lambda_{hhh}^{SM} = \frac{3 M_h^2}{M_Z^2}
\label{eq:3} \\
&MSSM:& \lambda_{hhh} = 3\cos(2\alpha) \sin(\beta+\alpha)
+ \frac{3 \epsilon}{M_Z^2} \frac{\cos^3\alpha}{\sin \beta} 
\end{eqnarray}
where the mixing angle $\alpha$ and $\beta$ are related by: 
\begin{equation}
\tan 2 \alpha=
 \frac{ M_{A}^2 + M_Z^2 }{ M_{A}^2 - M_Z^2 + \epsilon / \cos 2\beta} 
 \; \tan 2 \beta
\end{equation}
and the radiative corrections in the leading ${m_t}^4$ one-loop approximation, parametrized by \cite{Plehn:1996wb}
\begin{equation}
\epsilon = \frac{3 G_F}{\sqrt{2} \pi^2}
              \frac{m_t^4}{\sin^2 \beta} \;
              \log\left[ 1 + \frac{{M_S}^2}{m_t^2} \right]
\end{equation}

\begin{figure}[tb]
\begin{center}
{\includegraphics[angle =270, width=0.6\textwidth]{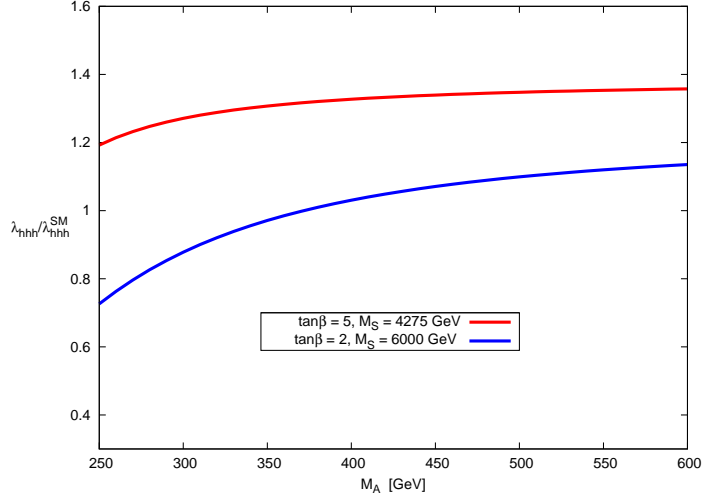} }
\caption{ {\it Variation of  ${\lambda_{hhh}}/{\lambda_{hhh}^{SM}}$ with $M_A$. 
Red (blue) line represents scenarios with $\tan\beta$ = 5 (2) and 
$M_S$ = 4275 GeV (6000 GeV). }}
\label{fig_coupling_ma}
\end{center}
\end{figure}

The Higgs pair production cross sections ($gg \rightarrow hh$) has 
dependence on the ${\lambda_{hhh}}$ which is determined by $M_A, \tan\beta$ 
and $M_S$. Di-Higgs production cross section can be  
enhanced/reduced from SM value, depending on the MSSM parameters.
For example, with ${\lambda_{hhh}}/{\lambda_{hhh}^{SM}}$ = 0, 1 and 2, 
the Higgs pair production cross sections  are 71 $fb$, 34 $fb$ and 16 $fb$ 
respectively  \cite{ATLAS-collaboration:1484890}. 
Cross section at the lower values of $\lambda_{hhh}$ increased 
due to the destructive interference of 
box and triangle diagrams involving $gg \rightarrow hh$. 

We show the variation of ${\lambda_{hhh}}/{\lambda_{hhh}^{SM}}$  as a function 
of $M_A$ for two sets of fixed tan$\beta$ and SUSY scale ($M_S$) in 
Fig. \ref{fig_coupling_ma}.  
The values of tan$\beta$ and $M_S$ are motivated from our selected benchmark 
points (see Table~\ref{tab2}). For tan$\beta$ = 2, the ratio is very close to 
unity for $M_A >$ 350 GeV and for tan$\beta$ = 5, the ratio varies from 1.2 to 1.35 depending 
on $M_A$. 



\section{Analysis and Results}
\label{sec:analysis}
At the leading order, SM Higgs pair production occurs either through gluon fusion to $hh$ (via top quark box diagram) 
or gluon gluon to virtual Higgs (mediated by top quark triangle diagram) and its splitting into a pair of Higgs bosons. 
The contribution of the box diagram is independent of the Higgs self coupling and this  also interfere destructively 
with the later one. At present, the SM Higgs pair cross section is known at the NLO level and it is about 34 fb 
at the 14 TeV LHC for $M_{h}=125$ GeV \cite{higgs_cs}. 
As discussed in the previous section, $hh$ production cross-section can be changed depending 
on MSSM parameters. However, for illustration purpose we only use the SM value. 
\\

Let us now discuss potentially detectable final states from Higgs pair production 
at the LHC. In spite of the large cross section of $b \bar b b \bar b$ final state, 
it is very difficult to observe the di-Higgs signal in this channel due to the huge 
QCD background. On the other hand, fully leptonic final states from $h$ to gauge boson decay, i.e., 
$h \rightarrow ZZ/WW \rightarrow $leptons, is not very promising due to small branching fractions of $W/Z$ 
 bosons to leptons. Recent studies using jet substructure technique show that 
 $b \bar{b} \tau^+ \tau^-$ \cite{Dolan:2012rv} and 
$b\bar b W^+W^-$ \cite{Papaefstathiou:2012qe} 
 channels may be encouraging 
at the HL-LHC. So far, the most promising channel to observe di-Higgs production is 
$b \bar b \gamma \gamma$ final state, although the branching to this particular 
channel is very small (about 0.27 \%). Observation of this channel is possible due 
to the high identification efficiency as well as excellent energy measurement of 
photons so that Higgs candidate in the di-photon invariant mass distribution can be easily 
separated from the background. The ATLAS collaboration has performed a 
detailed analysis \cite{ATLAS-collaboration:1484890} 
in this channel closely following the Ref.~\cite{Baur:2003gp} and according to their estimation 
di-Higgs ($hh$) production signal can be observed at $\sim$ $3\sigma$ level 
at the HL-LHC \footnote{The dominant SM background in this process is $t\bar{t}h$. It may be noted that we 
have only generated SUSY $H \rightarrow h h $ signal and SM $hh$ production processes and SM backgrounds are 
not analysed in our work.}
In this work, we only focus on the most promising final state, i.e., $b \bar b \gamma \gamma$ channel.\\

By comparing cross sections of single $H$ production (see Table~\ref{tab1}) and direct 
pair production of $h$, we can see that single $H$ cross section can be up to two orders 
in magnitude higher than the $hh$ cross section. 
Depending on the branching $H \rightarrow hh$, $H$ production can, in principle, 
contaminate the signal of direct $hh$ production and therefore 
affect the measurement of self coupling of 125 GeV Higgs (h). 
For illustration purpose, we take the benchmark points BP1 - BP5, 
introduced in Sec.~\ref{sec:benchmark} (see Table~\ref{tab2}) 
and compare with SM $hh$ production in the $b \bar{b} \gamma \gamma$ channel 
\footnote{All these benchmark points are well below the current LHC bound on heavy Higgs 
\cite{CMS-PAS-HIG-13-025,Aad:2014yja}.}.   \\

SM parton level $hh$ events with $M_h = 125$ GeV have been generated using MadGraph5 \cite{Alwall:2011uj} 
at the 14 TeV LHC using the model file of ``Higgs Pair Production" \cite{mad_higgs} which includes both 
top quark box and triangle diagrams. For generating 
MSSM signal (i.e., single $H$ production) we have used the MC generator PYTHIA \cite{Sjostrand:2006za}. 
All SM and SUSY events are showered and hadronized by PYTHIA and cross sections 
are scaled to NLO/NNLO values given in Table~\ref{tab2} for different benchmark 
points.  
For object reconstruction, we use fast detector simulator package 
Delphes3 \cite{deFavereau:2013fsa} \footnote{Jets are reconstructed with  anti-$k_t$ algorithm with $R = 0.4$.}. 
Details of the analysis cuts and efficiencies, following the ATLAS 
analysis \cite{ATLAS-collaboration:1484890}, are discussed below.\\

Events containing two $b$-jets and two photons are selected. Selection requirements 
of $b$-jets  are $p_T >$ 40 (25) GeV for leading (sub-leading jets) and  $|\eta|<$ 2.5. 
It is assumed that b-tagging efficiency is 80\%. Photons are selected with $p_T >$ 25 GeV,  
$|\eta|<$ 2.5 and they satisfy the isolation requirement. Following ATLAS analysis 
\cite{Aad:2010sp} photon candidate is removed if more than 4 GeV of transverse energy is 
observed within a cone with $\Delta R $= 0.4 surrounding the photon. The  photon 
identification efficiency is assumed to be 80\%. Separation cuts between $bb, b \gamma, 
\gamma \gamma$ pair ($\Delta R(b,b)$, $\Delta R(b,\gamma)$, $\Delta R(\gamma \gamma$)) 
are greater than 0.4. Criteria for reconstructing the Higgs mass are: $50$ GeV$<M_{b\bar b} <130$ GeV,  
$120$ GeV $<M_{\gamma\gamma} <130$ GeV, where $M_{b\bar b}$ and $M_{\gamma\gamma}$ are the invariant masses 
of $b \bar{b}$ and di-photon respectively. Finally, a lepton veto is also applied. After imposing the 
above cuts we obtain approximately 13.5 SM $hh$ events at the 14 TeV LHC with integrated 
luminosity $\cal L$ = 3000 $fb^{-1}$ which is fairly consistent with the ATLAS result 
\cite{ATLAS-collaboration:1484890}. 
On the other hand, the same cuts select 260, 779, 
13, 7 and 2 events 
for $BP1$, $BP2$, $BP3$, $BP4$, $BP5$ respectively with the same luminosity.
We find that the MSSM single $H$ production 
can change the di-Higgs production cross section and this effect is significant in the $H$ mass range 250-600 GeV 
in the low $\tan \beta$ region. \\

Counting of the number of $b \bar{b} \gamma \gamma$ events in all these benchmark 
points shows moderate to huge excess over SM cross section, although it is not possible 
to identify the origin of the excess from this counting itself. It is therefore important 
to separate out the direct $hh$ cross section from the MSSM heavy Higgs contribution. 
As the decay width of $H$ is small (at most a few GeV), one may try to separate out the direct $hh$ events 
from the MSSM $H$ events by identifying it in the $b \bar{b} \gamma \gamma$ invariant mass ($M_{hh}$)
distribution \cite{Dolan:2012ac,Chen:2013emb}.\\

\begin{figure}[!htb]
\begin{center}
\includegraphics[angle =270, width=0.75\textwidth]{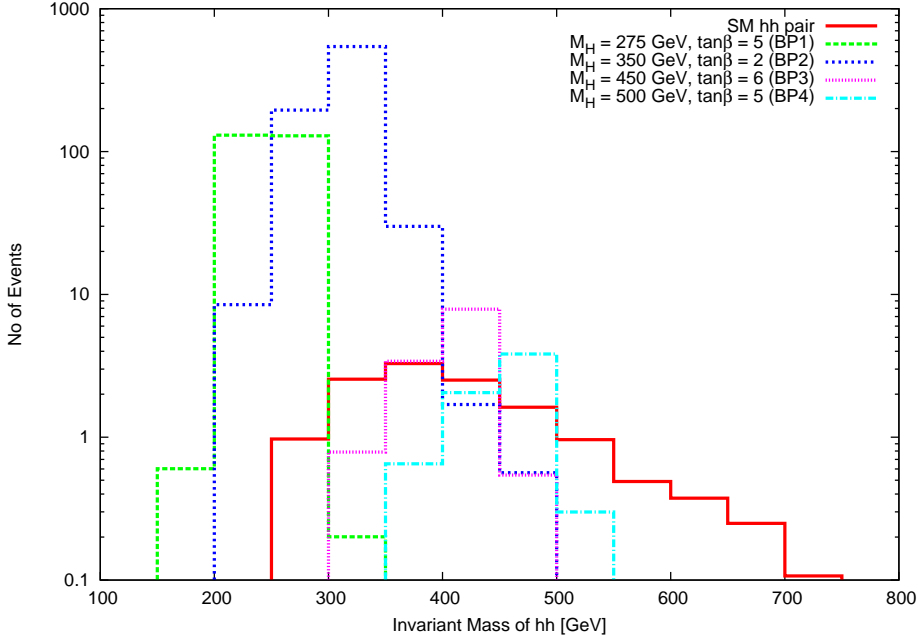} 
\caption{ {\it  Invariant mass distribution of SM $hh$ events and the events coming from 
single 
$H$ production for different SUSY benchmark points. Events 
are obtained after the basic cuts/Higgs reconstruction (see text) with $\cal L$ = 3000 $fb^{{-1}}$ 
at 14 TeV HL-LHC. }}
\label{fig:invmass}
\end{center}
\end{figure}

In Fig.\ref{fig:invmass} we present the invariant mass distribution of reconstructed $hh$ pair 
of SM events and SUSY events for four benchmark points (BP1-4)\footnote{The number of events for BP5 is very 
small and it is not shown in Fig.\ref{fig:invmass}. }. 
We can identify the clear peaks of heavy Higgs 
for $BP1$ and $BP2$. We can see that the $hh$ invariant mass distribution spreads over 100-150 GeV range, 
although, the $H$ decay width is small (about a few GeV).  This broadening  mainly comes from the reconstructed 
$h$ from the $b$ quark pair and this can not be reduced further because of the limitation of the hadronic 
calorimeter. It is therefore not possible to sharpen the $H$ mass peak in this channel. However, one can remove most 
of the $H$ events by putting a cut on the $hh$ invariant mass distribution around $H$ peak, although, the 
same cut may remove some amount of direct $hh$ events. Depending on the values of 
$M_H$ and $\tan \beta$ we may think of four possibilities:  \\

{\bf Scenario A:} MSSM contribution is very large compared to SM (for example, consider $BP1$). The excess 
events over direct $hh$ events can be separated by imposing a cut on di-Higgs invariant mass and in that case 
direct $hh$ pair events may be marginally affected by this cut. In case of $BP1$, one can remove almost 
all of the SUSY contribution (259 out of 260) by rejecting the events in the $M_{hh}$ bin of 200-300 GeV. 
However, this cut 
only reduces the SM events by one. Here we can separately measure both contributions and the determination 
of parameters in the Higgs sector may be possible in this case.   \\
 
{\bf Scenario B:}  In this case SUSY contribution is also large (for example, see  $BP2$) although invariant mass cut
may not help us to separate out these two contributions. The position of the $H$ peak is such that the events 
coming from the direct $hh$ production also large around $M_H$. In case of  $BP2$, there are 770 events (total events = 779) in the region 
$250$ GeV $<M_{hh}<400$ GeV and if we reject events of these bins, the SM $hh$ events reduces to 7 from 13.5. 
One possibility to 
separate these two contributions is to fit the $H$ peak and continuum $hh$ events simultaneously. However, the number of 
direct $hh$ events may not be statistically sufficient for this procedure. However, we can clearly identify existence of $H$  
from this measurement.   \\

{\bf Scenario C:} SUSY contribution is comparable or slightly smaller than SM in this case 
(for example, see $BP3$, $BP4$). Identification of distinct reconstructed peak of $H$ is difficult because of the poor statistics. The slight excess which comes from $H$, can be manifested as an enhancement of $h$ self coupling. However, many different new physics models can give 
rise to enhancement of $h$ self-coupling and it is thus difficult to identify the presence of heavy Higgs in this scenario. 
Invariant mass cut also do not help to identify the MSSM contribution in this case. 
 while it also removes 8 direct $hh$ events. 
Again a cut $350$ GeV$<M_{hh}<500$ GeV can remove 7 events of $BP4$ 
while it also removes about 7 direct $hh$ events. This should be regarded as the most 
challenging scenario in the context of the measurement of $h$ self coupling.  \\

{\bf Scenario D:} If we increase $M_A$ above 500-550 GeV, the number of events from direct $hh$ 
production reduces considerably and about 2-3 events survive. 
If $\tan\beta$ is in the range $\sim$ 5 or more (see BP5), H production cross section is 
also small and it is very difficult to observe the H resonance in that region. 
In case of BP5, we can expect 1-2 events which is not statistically significant.
For very small value of $\tan\beta$ ($<3$), there is a chance to observe the heavy Higgs boson 
resonance due to the enhancement of production cross section.

\section{Discussions}
\label{sec:conclusion}

In this paper we have studied how the decay of heavy MSSM Higgs can contribute the measurement of self 
coupling of lighter Higgs boson. The branching ratio $H\rightarrow hh$ can be sizable in the low $\tan\beta$ 
region and it can affect the direct $hh$ signal if the mass of H lies in between 250-600 GeV. 
Depending on the parameter space, 
MSSM signal can be very large compared to direct $hh$ production and in 
that case, clear identification of heavy Higgs boson is 
possible. However, we have identified a region in the parameter space corresponding 
to $M_H=400-600$ GeV with low $\tan \beta$ 
where the MSSM contribution is small but non-negligible. 
In such scenarios, the identification of $H$ is difficult and an excess in cross sections 
may be explained in terms of enhancement of $h$ self coupling.
This should be regarded as the most challenging scenario and further studies 
are required in this direction. 
Before concluding we are interested to point out a few relevant issues: 
\begin{itemize} 
\item In our work we rely on the default smearing parameters implemented in Delphes3. 
However, in case of HL-LHC, 
there can be some changes in the LHC detector design and one may expect more smearing 
effect compared to the conventional case. To check the effect of smearing we increase the 
default ATLAS smearing parameters of ECAL and HCAL by a factor of 25\% and study its 
effect. We find that our result is almost unaffected under this change.

\item One may think of discovering heavy Higgs bosons in different channel, for example $t\bar{t}$ final state which is the dominant 
decay mode for $M_H>350$ GeV. However, in that case, the main background is SM top quark production. Assuming that the reconstructed heavy Higgs mass may lie within 100 GeV mass range, our naive estimation shows that the ratio of SM $t\bar{t}$ background and signal can be as large as 100-1000 at the parton level, which makes it hard to detect in this mass range and it requires a dedicated study. It is also important to identify the other Higgs bosons $A$ and $H^{\pm}$ in this mass 
range and jet substructure technique can be useful in this purpose \cite{amit}.     

\item In this paper we consider heavy CP even Higgs production and its decay to 125 GeV Higgs in the $b\bar{b}\gamma\gamma$ channel. However, the same signal may come from other production processes like 
$pp \rightarrow A \rightarrow Zh, Z \rightarrow b \bar b, h \rightarrow \gamma \gamma$. In the Higgs self couping
measurement we accept $b \bar{b}$ invariant mass window in the range 50 GeV to 130 GeV and $Z \rightarrow b \bar{b}$ can easily contribute to the measurement.
However, Zh production can be seen in leptonic channels, for example, $\mu\mu \gamma \gamma$ etc. and this information can be used to separate out such contribution
from real $hh$ production.

\item Heavy Higgs production and its decay to $hh$ can interfere with direct $hh$ production 
process. This may be seen as dip/rise in the di-Higgs invariant mass distribution around 
$M_H$ mass at the parton level. However, we have checked that this effect is smeared out by 
hadronization and detector effects and it is difficult to identify the interference 
effect.

\item In this paper we only consider heavy Higgs to $hh$ decay. Depending on the parameter 
space, several potentially observable decay modes \cite{ATL-PHYS-PUB-2013-016,Craig:2013hca} may open 
and in principle, MSSM Higgs bosons can be discovered in other channels. 
A detailed study is beyond the scope of this paper and these issues will be considered in a 
separate work\cite{amit}.

\end{itemize}

\section{Acknowledgements}
We acknowledge Amit Chakraborty for useful discussion at the initial stage of the work.  
Work of BB is supported by Department of Science and Technology, Government
of INDIA under the Grant Agreement numbers IFA13-PH-75 (INSPIRE Faculty Award) and World Premier 
International Research Center Initiative, MEXT, Japan. BB also thanks Kavli IMPU(WPI) where a 
part of the work was done. The work of AC was partially supported by funding available 
from the Department of Atomic Energy, Government of India, for the Regional Centre for Accelerator-based
Particle Physics (RECAPP), Harish-Chandra Research Institute. 
AC also thanks IISER Kolkata where a part of the work was done. 

%


\begin{thebibliography}{99}
\markright{Bibliography}
\bibitem{Aad:2012tfa} 
  G.~Aad {\it et al.}  [ATLAS Collaboration],
  Phys.\ Lett.\ B {\bf 716}, 1 (2012)
[\href{http://arXiv.org/abs/arXiv:1207.7214}{arXiv:1207.7214}].


\bibitem{Chatrchyan:2012ufa} 
  S.~Chatrchyan {\it et al.}  [CMS Collaboration],
  Phys.\ Lett.\ B {\bf 716}, 30 (2012)
[\href{http://arXiv.org/abs/arXiv:1207.7235}{arXiv:1207.7235}].

\bibitem {ATLAS-CONF-2014-009} 
 ATLAS Collaboration,  ATLAS-CONF-2014-009 [\href{http://cds.cern.ch/record/1670012}{http://cds.cern.ch/record/1670012}].

\bibitem {CMS-PAS-HIG-14-009} 
 CMS Collaboration,  CMS-PAS-HIG-14-009 [\href{http://cds.cern.ch/record/1728249}{http://cds.cern.ch/record/1728249}]. 

\bibitem {ATLAS-collaboration:1484890} 
 ATLAS Collaboration,  ATL-PHYS-PUB-2012-004 [\href{http://cds.cern.ch/record/1484890}{http://cds.cern.ch/record/1484890}].

\bibitem{Boudjema:1995cb} 
  F.~Boudjema and E.~Chopin,
  Z.\ Phys.\ C {\bf 73}, 85 (1996) 
[\href{http://arXiv.org/abs/arXiv:hep-ph/9507396}{arXiv:hep-ph/9507396}].

\bibitem{Djouadi:1999rca} 
  A.~Djouadi, W.~Kilian, M.~Muhlleitner and P.~M.~Zerwas,
  Eur.\ Phys.\ J.\ C {\bf 10}, 45 (1999) 
[\href{http://arXiv.org/abs/arXiv:hep-ph/9904287}{arXiv:hep-ph/9904287}].

\bibitem{Plehn:1996wb} 
  T.~Plehn, M.~Spira and P.~M.~Zerwas,
  Nucl.\ Phys.\ B {\bf 479}, 46 (1996)
  [Erratum-ibid.\ B {\bf 531}, 655 (1998)] 
[\href{http://arXiv.org/abs/arXiv:hep-ph/9603205}{arXiv:hep-ph/9603205}].

\bibitem{Lafaye:2000ec} 
  R.~Lafaye, D.~J.~Miller, M.~Muhlleitner and S.~Moretti,
[\href{http://arXiv.org/abs/arXiv:hep-ph/0002238}{arXiv:hep-ph/0002238}].

\bibitem{Baur:2003gp} 
  U.~Baur, T.~Plehn and D.~L.~Rainwater,
  Phys.\ Rev.\ D {\bf 69}, 053004 (2004) 
[\href{http://arXiv.org/abs/arXiv:hep-ph/0310056}{arXiv:hep-ph/0310056}].

\bibitem{Dolan:2012rv} 
  M.~J.~Dolan, C.~Englert and M.~Spannowsky,
  JHEP {\bf 1210}, 112 (2012) 
[\href{http://arXiv.org/abs/arXiv:1206.5001}{arXiv:1206.5001}].

\bibitem{Baglio:2012np} 
  J.~Baglio, A.~Djouadi, R.~Gröber, M.~M.~Mühlleitner, J.~Quevillon and M.~Spira,
  JHEP {\bf 1304}, 151 (2013) 
[\href{http://arXiv.org/abs/arXiv:1212.5581}{arXiv:1212.5581}].

\bibitem{Goertz:2013kp} 
  F.~Goertz, A.~Papaefstathiou, L.~L.~Yang and J.~Zurita,
  JHEP {\bf 1306}, 016 (2013) 
[\href{http://arXiv.org/abs/arXiv:1301.3492}{arXiv:1301.3492}].

\bibitem{Yao:2013ika} 
  W.~Yao,
[\href{http://arXiv.org/abs/arXiv:1308.6302}{arXiv:1308.6302}].

\bibitem{Barger:2013jfa} 
  V.~Barger, L.~L.~Everett, C.~B.~Jackson and G.~Shaughnessy,
  Phys.\ Lett.\ B {\bf 728}, 433 (2014) 
[\href{http://arXiv.org/abs/arXiv:1311.2931}{arXiv:1311.2931}].


\bibitem{Dolan:2012ac} 
  M.~J.~Dolan, C.~Englert and M.~Spannowsky,
  Phys.\ Rev.\ D {\bf 87}, no. 5, 055002 (2013) 
[\href{http://arXiv.org/abs/arXiv:1210.8166}{arXiv:1210.8166}].

\bibitem{Chen:2013emb} 
  N.~Chen, C.~Du, Y.~Fang and L.~C.~Lü,
  Phys.\ Rev.\ D {\bf 89}, 115006 (2014) 
[\href{http://arXiv.org/abs/arXiv:1312.7212}{arXiv:1312.7212}].

\bibitem{Baglio:2014nea} 
  J.~Baglio, O.~Eberhardt, U.~Nierste and M.~Wiebusch,
  Phys.\ Rev.\ D {\bf 90}, 015008 (2014) 
[\href{http://arXiv.org/abs/arXiv:1403.1264}{arXiv:1403.1264}].

\bibitem{Hespel:2014sla} 
  B.~Hespel, D.~Lopez-Val and E.~Vryonidou,
[\href{http://arXiv.org/abs/arXiv:1407.0281}{arXiv:1407.0281}].

\bibitem{Osland:1998hv} 
  P.~Osland and P.~N.~Pandita,
  Phys.\ Rev.\ D {\bf 59}, 055013 (1999) 
[\href{http://arXiv.org/abs/arXiv:hep-ph/9806351}{arXiv:hep-ph/9806351}].


\bibitem{Djouadi:2006bz} 
  A.~Djouadi, M.~M.~Muhlleitner and M.~Spira,
  Acta Phys.\ Polon.\ B {\bf 38}, 635 (2007) 
[\href{http://arXiv.org/abs/arXiv:hep-ph/0609292}{arXiv:hep-ph/0609292}].

\bibitem{Han:2013sga} 
  C.~Han, X.~Ji, L.~Wu, P.~Wu and J.~M.~Yang,
  JHEP {\bf 1404}, 003 (2014) 
[\href{http://arXiv.org/abs/arXiv:1307.3790}{arXiv:1307.3790}].


\bibitem{Harlander:2012pb} 
  R.~V.~Harlander, S.~Liebler and H.~Mantler,
  Comput.\ Phys.\ Commun.\  {\bf 184}, 1605 (2013) 
[\href{http://arXiv.org/abs/arXiv:1212.3249}{arXiv:1212.3249}].

\bibitem{Martin:2009iq} 
  A.~D.~Martin, W.~J.~Stirling, R.~S.~Thorne and G.~Watt,
  Eur.\ Phys.\ J.\ C {\bf 63}, 189 (2009) 
[\href{http://arXiv.org/abs/arXiv:0901.0002}{arXiv:0901.0002}].
\bibitem{Martin:2009bu} 
  A.~D.~Martin, W.~J.~Stirling, R.~S.~Thorne and G.~Watt,
  Eur.\ Phys.\ J.\ C {\bf 64}, 653 (2009) 
[\href{http://arXiv.org/abs/arXiv:0905.3531}{arXiv:0905.3531}].

\bibitem{Martin:2010db} 
  A.~D.~Martin, W.~J.~Stirling, R.~S.~Thorne and G.~Watt,
  Eur.\ Phys.\ J.\ C {\bf 70}, 51 (2010) 
[\href{http://arXiv.org/abs/arXiv:1007.2624}{arXiv:1007.2624}].

\bibitem{Degrassi:2012ry} 
  G.~Degrassi, S.~Di Vita, J.~Elias-Miro, J.~R.~Espinosa, G.~F.~Giudice, G.~Isidori and A.~Strumia,
  JHEP {\bf 1208}, 098 (2012) 
[\href{http://arXiv.org/abs/arXiv:1205.6497}{arXiv:1205.6497}].

\bibitem{Djouadi:2013vqa} 
  A.~Djouadi and J.~Quevillon,
  JHEP {\bf 1310}, 028 (2013) 
[\href{http://arXiv.org/abs/arXiv:1304.1787}{arXiv:1304.1787}].


\bibitem{higgs_cs} 
[\href{https://twiki.cern.ch/twiki/bin/view/LHCPhysics/HiggsEuropeanStrategy2012}{https://twiki.cern.ch/twiki/bin/view/LHCPhysics/HiggsEuropeanStrategy2012}], 
M. Spira, 
[\href{http://tiger.web.psi.ch/hpair/}{http://tiger.web.psi.ch/hpair/}] (2012). 




\bibitem{atlas_gamma} 
  G.~Aad {\it et al.}  [ATLAS Collaboration],
  Phys.\ Rev.\ D {\bf 90}, 112015 (2014) 
[\href{http://arXiv.org/abs/arXiv:1408.7084}{arXiv:1408.7084}].


\bibitem{cms_gamma} 
  V.~Khachatryan {\it et al.}  [CMS Collaboration],
  Eur.\ Phys.\ J.\ C {\bf 74}, no. 10, 3076 (2014) 
[\href{http://arXiv.org/abs/arXiv:1407.0558}{arXiv:1407.0558}].

\bibitem{atlas_zz} 
  G.~Aad {\it et al.}  [ATLAS Collaboration],
[\href{http://arXiv.org/abs/arXiv:1408.5191}{arXiv:1408.5191}].


\bibitem{cms_zz} 
  S.~Chatrchyan {\it et al.}  [CMS Collaboration],
  Phys.\ Rev.\ D {\bf 89}, 092007 (2014) 
[\href{http://arXiv.org/abs/arXiv:1312.5353}{arXiv:1312.5353}].


\bibitem{atlas_ww} 
  G.~Aad {\it et al.}  [ATLAS Collaboration],
[\href{http://arXiv.org/abs/arXiv:1412.2641}{arXiv:1412.2641}].


\bibitem{cms_ww} 
  S.~Chatrchyan {\it et al.}  [CMS Collaboration],
  JHEP {\bf 1401}, 096 (2014) 
[\href{http://arXiv.org/abs/arXiv:1312.1129}{arXiv:1312.1129}].

\bibitem{atlas_bb} 
  G.~Aad {\it et al.}  [ATLAS Collaboration],
[\href{http://arXiv.org/abs/arXiv:1409.6212}{arXiv:1409.6212}].


\bibitem{cms_bb} 
  S.~Chatrchyan {\it et al.}  [CMS Collaboration],
  Phys.\ Rev.\ D {\bf 89}, no. 1, 012003 (2014) 
[\href{http://arXiv.org/abs/arXiv:1310.3687}{arXiv:1310.3687}].

\bibitem{atlas_tau} 
 ATLAS Collaboration,  ATLAS-CONF-2014-061 [\href{http://cds.cern.ch/record/1954724}{http://cds.cern.ch/record/1954724}].

\bibitem{cms_tau} 
  S.~Chatrchyan {\it et al.}  [CMS Collaboration],
  JHEP {\bf 1405}, 104 (2014) 
[\href{http://arXiv.org/abs/arXiv:1401.5041}{arXiv:1401.5041}].

\bibitem{Djouadi_rev} 
  A.~Djouadi, 
  Phys.\ Rept.\  {\bf 459}, 1 (2008) 
[\href{http://arXiv.org/abs/arXiv:hep-ph/0503173}{arXiv:hep-ph/0503173}].



\bibitem{Papaefstathiou:2012qe} 
  A.~Papaefstathiou, L.~L.~Yang and J.~Zurita,
  Phys.\ Rev.\ D {\bf 87}, 011301 (2013) 
[\href{http://arXiv.org/abs/arXiv:1209.1489}{arXiv:1209.1489}].

\bibitem {CMS-PAS-HIG-13-025}
  CMS Collaboration, CMS-PAS-HIG-13-025, [\href{http://cds.cern.ch/record/1637776}{http://cds.cern.ch/record/1637776}].

\bibitem{Aad:2014yja} 
  G.~Aad {\it et al.}  [ATLAS Collaboration],
[\href{http://arXiv.org/abs/arXiv:1406.5053}{arXiv:1406.5053}]. 


\bibitem{Alwall:2011uj} 
  J.~Alwall, M.~Herquet, F.~Maltoni, O.~Mattelaer and T.~Stelzer,
  JHEP {\bf 1106}, 128 (2011) 
[\href{http://arXiv.org/abs/arXiv:1106.0522}{arXiv:1106.0522}].

\bibitem{mad_higgs}
  \href {https://cp3.irmp.ucl.ac.be/projects/madgraph/wiki/HiggsPairProduction}
  {https://cp3.irmp.ucl.ac.be/projects/madgraph/wiki/higgspairproduction}.

\bibitem{Sjostrand:2006za} 
  T.~Sjostrand, S.~Mrenna and P.~Z.~Skands,
  JHEP {\bf 0605}, 026 (2006) 
[\href{http://arXiv.org/abs/arXiv:hep-ph/0603175}{arXiv:hep-ph/0603175}].


\bibitem{deFavereau:2013fsa} 
  J.~de Favereau {\it et al.}  [DELPHES 3 Collaboration],
  JHEP {\bf 1402}, 057 (2014) 
[\href{http://arXiv.org/abs/arXiv:1307.6346}{arXiv:1307.6346}].

\bibitem{Aad:2010sp} 
  G.~Aad {\it et al.}  [ATLAS Collaboration],
  Phys.\ Rev.\ D {\bf 83}, 052005 (2011) 
[\href{http://arXiv.org/abs/arXiv:1012.4389}{arXiv:1012.4389}].

\bibitem {amit} B. Bhattacherjee, A. Chakraborty, and A. Choudhury, (2014), 
[\href{http://arXiv.org/abs/arXiv:XXXX.XXXX}{(In Preparation)}].

\bibitem{ATL-PHYS-PUB-2013-016} 
  ATLAS Collaboration, ATL-PHYS-PUB-2013-016, [\href{http://cds.cern.ch/record/1611190}{http://cds.cern.ch/record/1611190}].

\bibitem{Craig:2013hca} 
  N.~Craig, J.~Galloway and S.~Thomas,
[\href{http://arXiv.org/abs/arXiv:1305.2424}{arXiv:1305.2424}].



\end{thebibliography}
\end{document}